\documentclass{webofc}

\usepackage[varg]{txfonts}   
\usepackage{hyperref}
\usepackage{url}
\usepackage{fancyhdr}
\hypersetup{colorlinks=true,citecolor=blue,urlcolor=blue,linkcolor=blue}
\begin{document}
\title{Measurements of Kaon Femtoscopy in Au+Au Collisions at $\sqrt{s_{NN}}$ = 3.0 - 4.5 GeV by the STAR Experiment}
%
%

\author{\firstname{Bijun} \lastname{Fan}\inst{1}\fnsep\thanks{\email{bjfan@mails.ccnu.edu.cn}}
(for the STAR collaboration)}

\institute{Central China Normal University, Wuhan 430079, China}

\abstract{In these proceedings, we present the measurements of charged $K^{+} - K^{+}$ and neutral $K_{s}^{0} - K_{s}^{0}$ correlation functions from Au+Au fixed-target collisions at $\sqrt{s_{NN}}$ = 3.0, 3.2, 3.5, 3.9 and 4.5 GeV at STAR. This is the first such systematic measurement of correlation functions involving strangeness in the high baryon density region. The source size values do not exhibit a clear energy dependence, and the transverse mass dependence of source size for kaons does not align with the trend observed for pions. Parameters extracted from UrQMD transport model calculations qualitatively capture the measured values. 
}
\maketitle
\section{Introduction}
\label{intro}
In heavy-ion collisions, the moment when the last scattering occurs among hadrons is referred to as "kinetic-freeze-out". Femtoscopy is a technique used to unravel the space-time and dynamical characteristics of the particle-emitting source at kinetic freeze-out. Kaons, in comparison to pions, experience reduced contributions from long-lived resonance decays and exhibit smaller rescattering cross-sections with hadronic matter, offering a more direct insight into the source. Theoretical momentum correlations between two identical particles are defined as the first part:
\begin{equation}
\begin{split}
    &C(p_{1}, p_{2}) = \int S(\vec{r^*})|\Psi(q_{\text{inv}},\vec{r^*})|^2 \, \mathrm{d}^3 \vec{r^*} = \frac{\mathcal{N}(N_{\text{same}}(q_{\text{inv}}))}{N_{\text{mixed}}(q_{\text{inv}})}
    \label{eq:Formula CF}
\end{split}
\end{equation}
where the $S(\vec{r^*})$ is emission function and $\Psi(q_{\text{inv}},\vec{r^*})$ is the two-particle wave function, $q_{\text{inv}}=\sqrt{(\vec{p_1}-\vec{p_2})^2 - (E_1-E_2)^2}$ means relative invariant momentum of the pair, and $\vec{r^*}$ represents relative distance of the emitters in the Pair Rest Frame(PRF). The experimental method to calculate the correlation function can be presented as the second part, in which $N_{same}(q_{inv})$ and $N_{mixed}(q_{inv})$ is the distribution of $q_{inv}$ formed by the same event and separate events but with same multiplicity and position of primary vertex, the factor $\mathcal{N}$ used to normalize the correlation function to unity at large relative momentum.
For charged kaons and pions, Sinyukov-Bowler \cite{coulomb} method is used to parameterize the correlation function 
\begin{equation}
    C(q_{inv}) = N[(1 - \lambda) + \lambda K_{Coul}(q_{inv},R_{G})(1+e^{-R_{G}^{2}q_{inv}^{2}})]
    \label{eq:coulCF}
\end{equation}
where $R_{G}$ is the source size, $\lambda$ is correlation strength, and $K_{Coul}(q_{inv},R_{G})$ is Coulomb factor.
For neutral kaons, it is Lednicky-Lyuboshitz \cite{llfit} approach shown as,
\begin{equation}
    C(q_{inv})=1 + e^{-q_{inv}^{2}R_{G}^{2}} + \frac{1 - \epsilon^{2}}{2}\left[\left|\frac{f(k^{*})}{R_{G}}\right|^{2} + \frac{4\mathcal{R}f(k^{*})}{\sqrt{\pi}R_{G}}F_{1}(q_{inv}R_{G}) - \frac{2 \mathcal{I} f(k^{*})}{R_{G}}F_{2}(q_{inv}R_{G})\right]
    \label{eq:K0sfitfunc}
\end{equation}
where $F_{1}(z)=\int _{0}^{z}dx \frac{e^{x^{2} - z^{2}}}{z}$ and $F_{2} = \frac{1 - e^{-z^{2}}}{z}$.
Here, the $\epsilon$ is the kaon abundance asymmetry, it can be calculate by yield of kaons: $\epsilon = \frac{K - \Bar{K}}{K + \Bar{K}}$, and $f(k^{*})$ is the $s$-wave scattering amplitude \cite{llfkstar}, where $k^{*}=\frac{1}{2}q_{inv}$ for identcal particles. 

\section{Data sets and Analysis strategy}

The $\sqrt{s_{NN}}$ = 3.0 - 4.5 GeV Au + Au collision data were collected by STAR. The primary vertex position of each event along the beam direction, $V_{z}$, is limited to be 198 cm $< V_{z} <$ 202 cm from the center of TPC (Time Projection Chamber). To eliminate possible beam interactions with the vacuum pipe, the vertex along the radial direction, $V_{r}$, is selected to be smaller than 2 cm (for $\pi^{+}$ and $K^{+}$) and 1.5 cm (for $K_{s}^{0}$). TPC and TOF (Time of flight) were used to identify the $\pi^{+}$ and $K^{+}$, and $K_{s}^{0}$ were reconstructed by Kalman Filter \cite{kfparicle} Particle package via $\pi^{+}\pi^{-}$ chanel.

\section{Results and Discussions}

\begin{figure}[!htbp]
\centering
\includegraphics[width=0.81\textwidth]{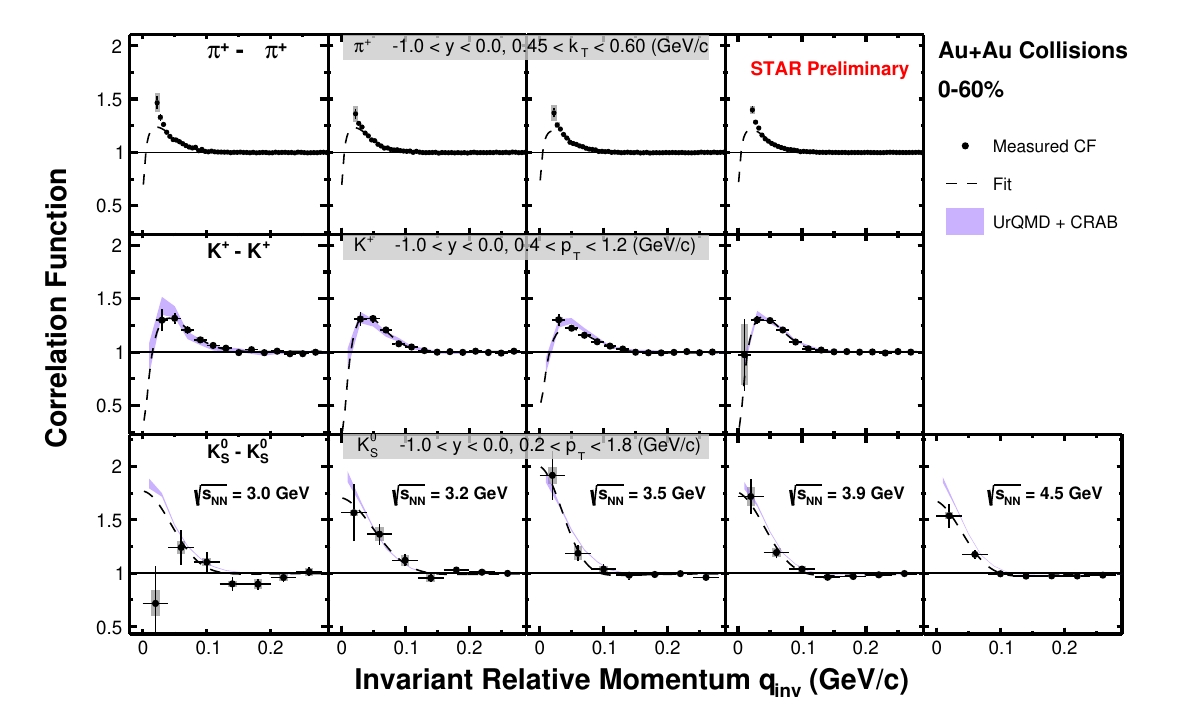}
\caption{One dimensional correlation functions of $\pi^{+} - \pi^{+}$ (upper panel), $K^{+} - K^{+}$  (middle panel), $K_{s}^{0} - K_{s}^{0}$ (bottom panel) in 0-60\% Au + Au collisions at $\sqrt{s_{NN}}$ = 3.0 GeV - 4.5 GeV. Vertical lines and grey vertical band represent the statistical and systematic uncertainties, respectively.}
\label{fig:CF}
\end{figure}
Figure~\ref{fig:CF} shows the one dimensional correlation functions of $\pi^{+} - \pi^{+}$, $K^{+} - K^{+}$  and $K_{s}^{0} - K_{s}^{0}$ from $\sqrt{s_{NN}}$ = 3.0 - 4.5 GeV Au + Au collisions at 0-60\% centrality. The purple bands are correlation functions calculated from transport model UrQMD \cite{urqmd} plus CRAB (Correlation Afterburner) \cite{crab}.

\begin{figure}[!htbp]
\centering
\includegraphics[width=0.84\textwidth]{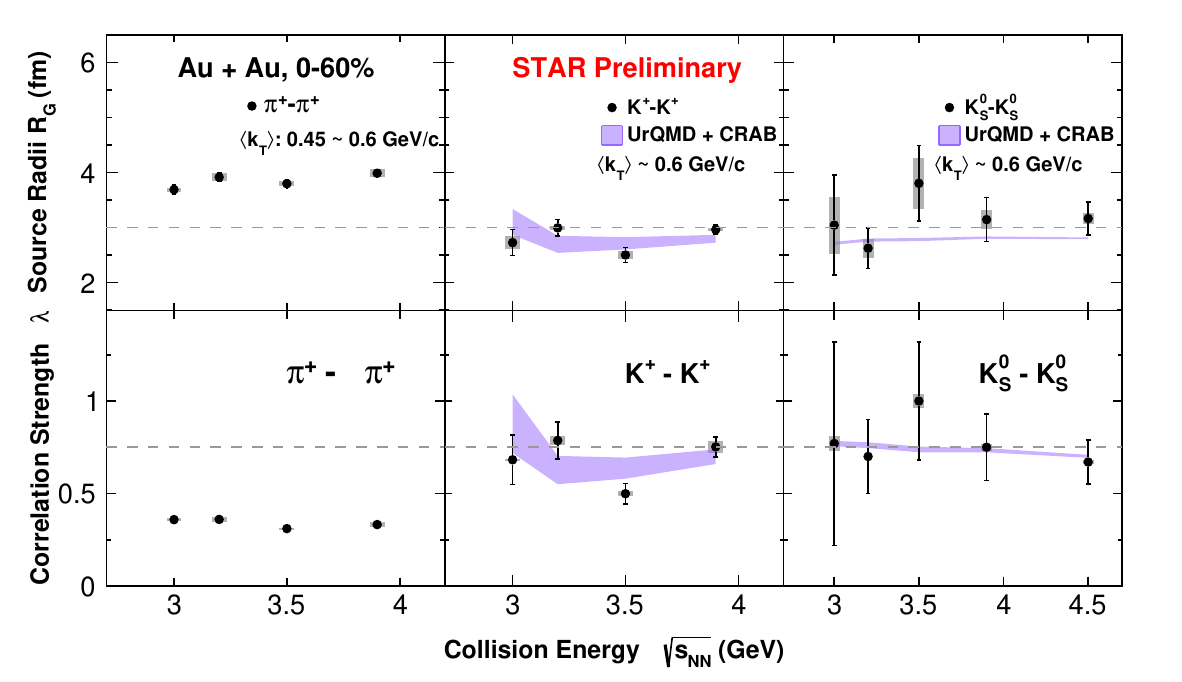}
\caption{Energy dependence of extracted parameters of $\pi^{+} - \pi^{+}$ (left panel), $K^{+} - K^{+}$  (middle panel), $K_{s}^{0} - K_{s}^{0}$ (right panel) in 0-60\% Au + Au collisions at $\sqrt{s_{NN}}$ = 3.0 - 4.5 GeV.}
\label{fig:ParCF}
\end{figure}

Figure~\ref{fig:ParCF} shows the energy dependence of extracted $R_{G}$ and  $\lambda$ of $\pi^{+} - \pi^{+}$, $K^{+} - K^{+}$ , $K_{s}^{0} - K_{s}^{0}$ pairs in Au + Au 0-60\% collisions at $\sqrt{s_{NN}}$ = 3.0 - 4.5 GeV. The grey vertical band represents systematical uncertainties, while the purple band is the result from UrQMD including CRAB to account for femtoscopic correlations, they are consistent with the data within uncertainty. There is no clear energy dependence of source radii observed in this energy region. The $\lambda$ parameter of pions is smaller than that of kaons, reflecting the fact that kaons experience less contribution from resonance decays. The $R_{G}$ and $\lambda$ parameters of charged kaons are consistent with those of neutral kaons within the uncertainties.

\begin{figure}[!htbp]
\centering
\includegraphics[width=0.4\textwidth]{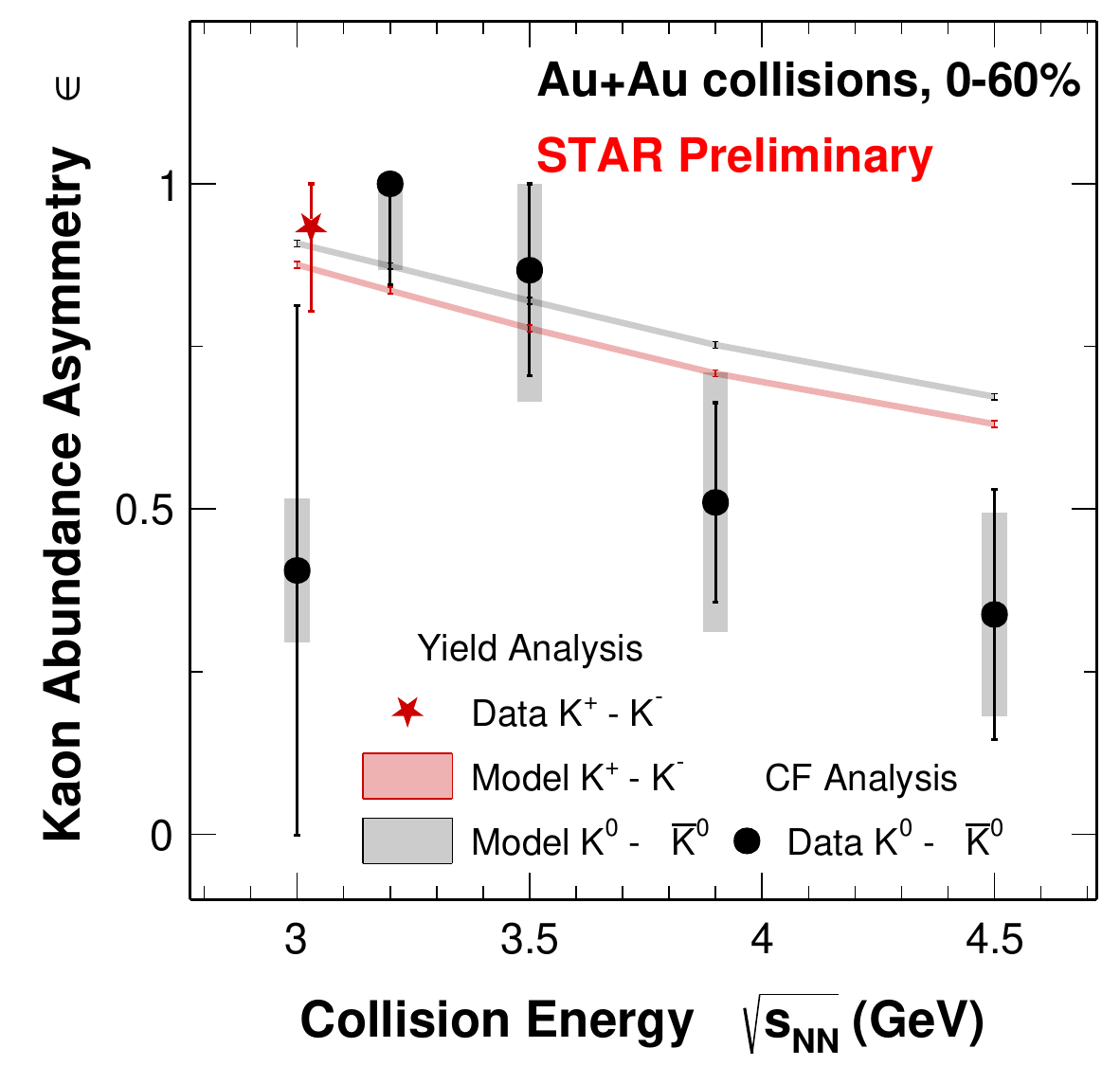}
\caption{Energy dependence of kaon abundance asymmetry of charged kaons (red point) and neutral kaons (black points) in Au + Au collisions at $\sqrt{s_{NN}}$ = 3.0 GeV - 4.5 GeV.}
\label{fig:Epsilon}
\end{figure}

The $K_{s}^{0} - K_{s}^{0}$ correlation function provides insights into the asymmetry of neutral kaon abundance, as $K_{s}^{0}$  is a mixture of $s$ and $\bar{s}$ quarks. In Fig. \ref{fig:Epsilon}, the black solid points represent the neutral kaon abundance asymmetry extracted from $K_{s}^{0} - K_{s}^{0}$ correlation functions, while the red point shows the charged kaon abundance asymmetry obtained from yield analysis based on STAR preliminary particle ratio results \cite{Benjamin}. The UrQMD calculations are depicted as black and red bands for neutral and charged kaons, respectively. These calculations are derived from yield analysis of the model. This study marks the first measurement of neutral kaon abundance asymmetry in heavy-ion collisions. The results indicate a decrease in abundance asymmetry with increasing collision energy. The model results align with the data trend, suggesting that associated production dominates in the high baryon density region, while pair production becomes more significant at higher collision energies.

\begin{figure}[!htbp]
\centering
\includegraphics[width=0.4\textwidth]{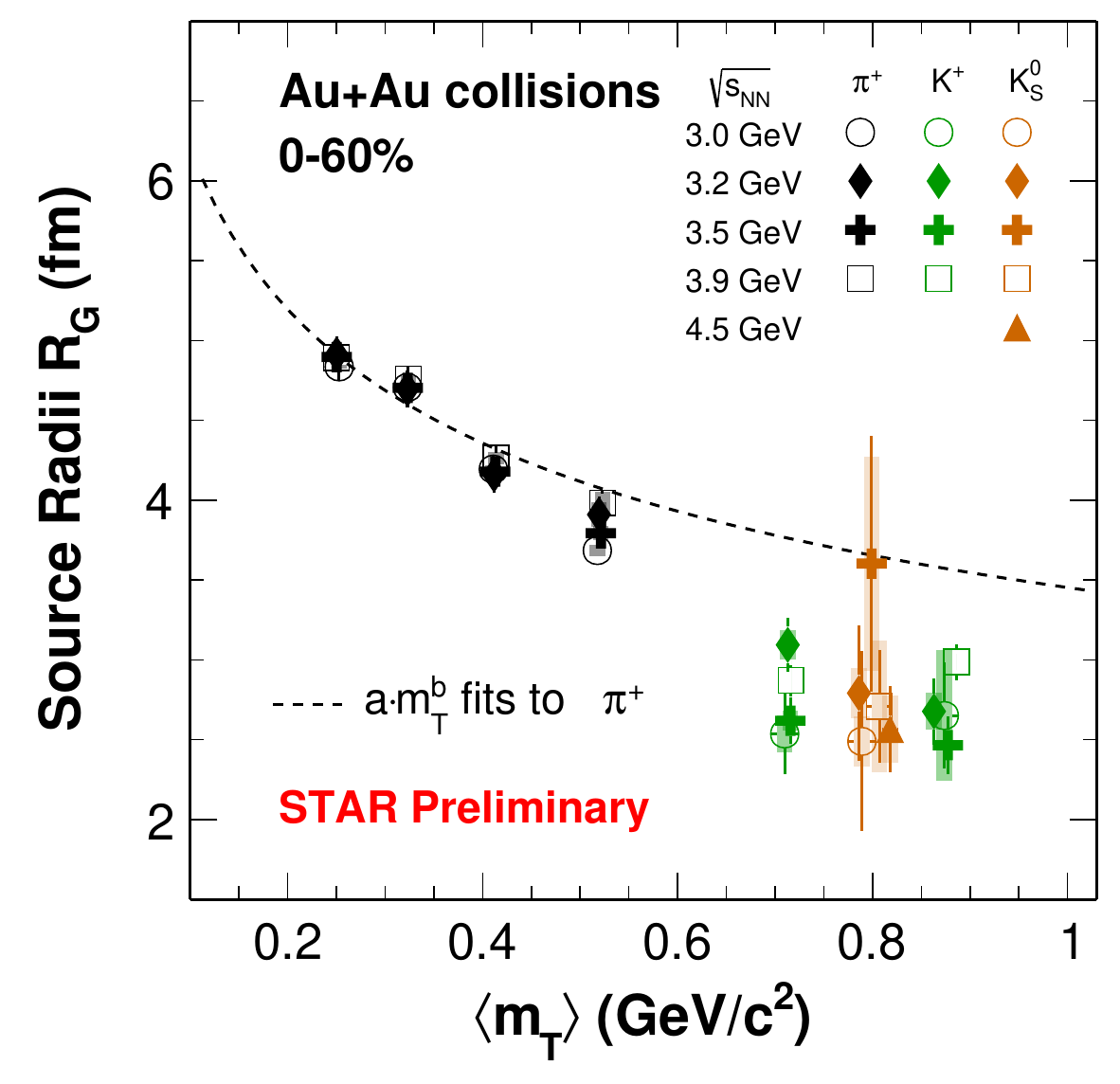}
\caption{The average transverse mass dependence of source radii $R_{G}$ of $\pi^{+} - \pi^{+}$, $K^{+} - K^{+}$ , $K_{s}^{0} - K_{s}^{0}$ in 0-60\% Au + Au collisions at $\sqrt{s_{NN}}$ = 3.0 GeV - 3.9 GeV. The $R_{G}$ of $\pi^{+} - \pi^{+}$ is represented by black points, while the $K^{+} - K^{+}$  and $K_{s}^{0} - K_{s}^{0}$ are represented by green and brown points, respectively. The black dashed line represents a power-law fit to all the data points for pions, with the power-law exponent b is -0.25.}
\label{fig:mtscaling}
\end{figure}

The average transverse mass dependence of source radii can also be examined. In Fig.~ \ref{fig:mtscaling}, the black dashed line corresponds to the power-law fitting of all data points from pions. The radii of pions are expected to follow the same scaling due to the universal collective flow predicted by hydrodynamics \cite{mtcolloctive} at higher energies, while the $R_{G}$ of kaons does not follow the trend of pions, suggesting a lack of equilibrium between pions and kaons in this energy range.

\section{Summary}
We reported the first measurements of kaon femtoscopy in Au+Au collisions in the high baryon density region, and the source parameters are extracted. However, $R_{G}$ parameter of kaons does not follow the $m_{T}$ scaling from pion correlation functions which may indicate there is no equilibrium amongst pions and kaons at these collisions. Meanwhile, the abundance asymmetry parameter of neutral kaon is determined and they are decreasing as a function of the collision energy, it indicates that the pair production becomes more important at higher collision energies. The UrQMD model reproduced all the above observations for kaons within uncertainties.\\

\noindent \textbf{Acknowledgements} This work is supported in part by the National Key Research and Development Program of China under Contract No. 2022YFA1604900; the National Natural Science Foundation of China (NSFC) under contract No. 12175084.
%
%
%

\end{document}